# Memory effects in black phosphorus field effect transistors

*Alessandro Grillo, Aniello Pelella, Enver Faella, Filippo Giubileo, Stephan Sleziona, Osamah Kharsah, Marika Schleberger, and Antonio Di Bartolomeo*[*]

Alessandro Grillo, Author 1,
Physics Department "E. R. Caianiello" and Interdepartmental Center "Nanomates",
University of Salerno, via Giovanni Paolo II n. 132, Fisciano 84084, Italy
E-mail: agrillo@unisa.it

Aniello Pelella, Author 2,
Physics Department "E. R. Caianiello" and Interdepartmental Center "Nanomates",
University of Salerno, via Giovanni Paolo II n. 132, Fisciano 84084, Italy
E-mail: apelella@unisa.it

Enver Faella, Author 3,
Physics Department "E. R. Caianiello" and Interdepartmental Center "Nanomates",
University of Salerno, via Giovanni Paolo II n. 132, Fisciano 84084, Italy
E-mail: efaella@unisa.it

Dr. Filippo Giubileo, Author 4,
CNR-SPIN, via Giovanni Paolo II n. 132, Fisciano 84084, Italy
E-mail: filippo.giubileo@spin.cnr.it

Dr. Stephan Sleziona, Author 5,
Fakultät für Physik and CENIDE, Universität Duisburg-Essen, Lotharstrasse 1, Duisburg D-47057, Germany
E-mail: stephan.sleziona@uni-due.de

Osamah Kharsah, Author 6,
Fakultät für Physik and CENIDE, Universität Duisburg-Essen, Lotharstrasse 1, Duisburg D-47057, Germany
E-mail: osamah.kharsah@stud.uni-due.de

Prof. Marika Schleberger, Author 7,
Fakultät für Physik and CENIDE, Universität Duisburg-Essen, Lotharstrasse 1, Duisburg D-47057, Germany
E-mail: marika.schleberger@uni-due.de

Prof. Antonio Di Bartolomeo, Author 8,
Physics Department "E. R. Caianiello" and Interdepartmental Center "Nanomates",
University of Salerno, via Giovanni Paolo II n. 132, Fisciano 84084, Italy
E-mail: adibartolomeo@unisa.it

We report the fabrication and the electrical characterization of back-gated field effect transistors with black phosphorus channel. We show that the hysteresis of the transfer characteristic, due to intrinsic defects, can be exploited to realize non-volatile memories. We demonstrate that gate voltage pulses allow to trap and store charge inside the defect states, which enable memory devices with endurance over 200 cycles and retention longer than 30 minutes. We show that the use of a protective poly (methyl methacrylate) layer, positioned on top of the black phosphorus channel, does not affect the electrical properties of the device but avoids the degradation caused by the exposure to air.

## 1. Introduction

Black phosphorus (BP) is a layered material in which, similarly to graphite, individual atomic layers are held together by van der Waals interactions.[1] In the single-layer limit, BP is also known as phosphorene and has a numerically predicted direct band gap of $\sim 2\ eV$ at the $\Gamma$ point of the first Brillouin zone.[2] With the increasing number of layers, the interlayer interactions reduce the bandgap to a minimum of $0.3\ eV$ for bulk BP[3] moving the direct gap to the Z point.[4]

Preferably, BP presents an orthorhombic crystal structure with puckered layers[1] in which each phosphorus atom is connected with the nearest three atoms through covalent bonds, thus forming layers with a corrugated shape. Theoretical studies have indicated that BP could have different structures such as simple cubic, diamond, orthogonal and nanorod, but the orthorhombic structure is the thermodynamically most stable allotrope at 0 K.[5] The distance between two neighbouring layers is approximately $0.5\ nm$, slightly larger than the interlayer distance in graphite ($0.36\ nm$). Hence, BP can store charged ions better than graphite and has been proposed for energy storage applications.[6] Similarly to transition metal dichalcogenides[7–14], BP's band structure has attracted a lot of attention for possible electronic and optoelectronic applications. Indeed, the presence of a finite bandgap makes BP suitable for the realization of field-effect transistors (FETs), and the thickness-dependent direct bandgap may lead to applications in optoelectronics, especially in the infrared region. Several devices have already been proposed and studied; Li et al.[15] achieved reliable transistor performance with five orders of magnitude drain current modulation and charge-carrier mobility up to $\sim 1000\ cm^2\ V^{-1}\ s^{-1}$ obtained from $10\ nm$ thick samples. Ren et al.[16] used BP nanosheets to realize self-powered photodetectors with superior photoresponse activity under light irradiation and environmental robustness, showing that BP is a promising building block for optoelectronic devices. Moreover, BP has been also proposed for solar cells,[17] as humidity sensor,[18] or for live cell imaging[19].

A remarkable application of BP has been in the field of memory devices. Common non-volatile FET-based memories use a charge-trapping layer to accumulate and retain the electric charge induced by a gate pulse. Similar devices, with channel based on several 2D-materials, including BP, have been already proposed and demonstrated. In particular, Feng et al.[20] reported that few-layer BP channel FETs with a $Al_2O_3/HfO_2/Al_2O_3$ charge-trapping gate stack enable memory devices with programming window exceeding $12\ V$, due to the extraordinary trapping ability of high-k $HfO_2$, and endurance exceeding 120 cycles. Tian et al.[21] used a similar structure, where the charge trapping layer consisted only of $Al_2O_3$, to demonstrate an ambipolar BP charge-trap memory device with dynamically reconfigurable and polarity-reversible memory behaviour. In such a device, the polarity of the carriers in the BP channel can be reversibly switched between electron- and hole-dominated conductions allowing four different memory states and, hence, two-bit per cell data. Finally, Lee et al.[22] showed that gold nanoparticles as charge trapping layer for mechanically-exfoliated few-layered BP FETs enable large memory window ($58.2\ V$), stable retention ($10^4\ s$), and cyclic endurance (1000 cycles).

However, the practical implementation of these promising device is hindered by the intrinsic instability of mono- and few-layer BP. While bulk crystals are quite stable in air, BP flakes thinner than $10\ nm$ degrade in few days, or even in few hours if the thickness is reduced to the single layer.[1] Indeed, when exposed to oxygen, devices obtained through mechanical exfoliation[23] or solvent exfoliation[24] of bulk crystals are affected by fast oxidative degradation.[25,26] Moreover, the degradation process can be enhanced by exposure to light through photo-oxidation.[27] To tackle with this issue, different stabilization processes have been tried. As first attempt, BP has been functionalized through $TiO_2$, $Al_2O_3$, titanium sulfonate ligand ($TiL_4$), polyimide, or aryl diazonium.[28–32] Encapsulation with other 2D materials, such as graphene and boron nitride, has been considered as well.[33] Furthermore, it has been reported that the use of ionic liquids can suppress BP degradations for months.[34] Although these first approaches have provided encouraging results, an efficient stabilization of BP, also compatible with standard fabrication processes of high performance devices remains still an ongoing challenge. In this work, we report the fabrication and the electrical characterization of back-gated BP FETs. We exploited the presence of intrinsic and BP/$SiO_2$ interface defects to realize non-volatile memories with good endurance and retention properties. By comparing samples exposed directly to the air with others covered by poly (methyl methacrylate) (PMMA), we demonstrate that PMMA does not affect the electrical properties of the devices but prevents the channel degradation at least for one month.

## 2. Materials and methods

Figure 1a shows the crystal structure of BP where the layered structure is composed of sheets with the phosphorus atoms arranged in a puckered honeycomb lattice. Ultrathin BP flakes were exfoliated from bulk BP single crystals (Smart Elements) using a standard mechanical exfoliation method by adhesive tape. The flakes were transferred onto degenerately doped p-type silicon substrates, covered by $90\ nm$-thick $SiO_2$, on which they were located through optical microscopy. A standard photolithography process followed by electron beam evaporation was used to deposit $10\ nm\ Cr/100\ nm\ Au$ electrodes, as shown in Figure 1b. A back-gate contact was formed covering the scratched area of the Si substrate with silver paste. Electrical measurements were carried out in two- and four-probe configurations in a Janis ST-500 Probe Station equipped with four nanoprobes connected to a Keithley 4200 SCS (semiconductor characterization system), working as source-measurement unit with current and voltage sensitivity better than 1 pA and 2 $\mu V$, respectively. The electrical characterization was performed at room temperature and in high vacuum at a pressure of $\sim 10^{-5}\ mbar$.

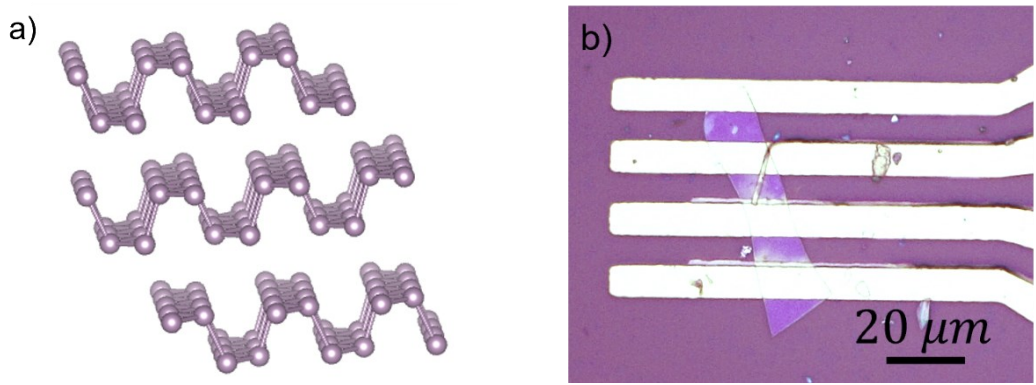

Figure 1 – a) Crystal structure of few-layer BP. b) Optical image of the device showing a BP flake covered by four $Cr/Au$ contacts.

## 3. Results and discussion

Figure 2a reports a schematic of the device showing a standard four-probe characterization set-up. The current ($I_{xx}$) is forced between the outer contacts while the voltage drop ($V_{ds}$) is measured between the inner contacts. Figure 2b shows the characteristics for both 4-probe ($I_{xx} - V_{ds}$) and 2-probe ($I_{ds} - V_{ds}$) configurations resulting in a channel resistance of 3.5 $k\Omega$ and 3.6 $k\Omega$, respectively. Since the difference between the two methods is less than 3%, indicating that the contacts are ohmic with low contact resistance[35,36], in the following we use the 2-probe configuration for easier measurements. The electrical characterization of the BP transistor at room temperature is reported in Figure 2c and 2d.

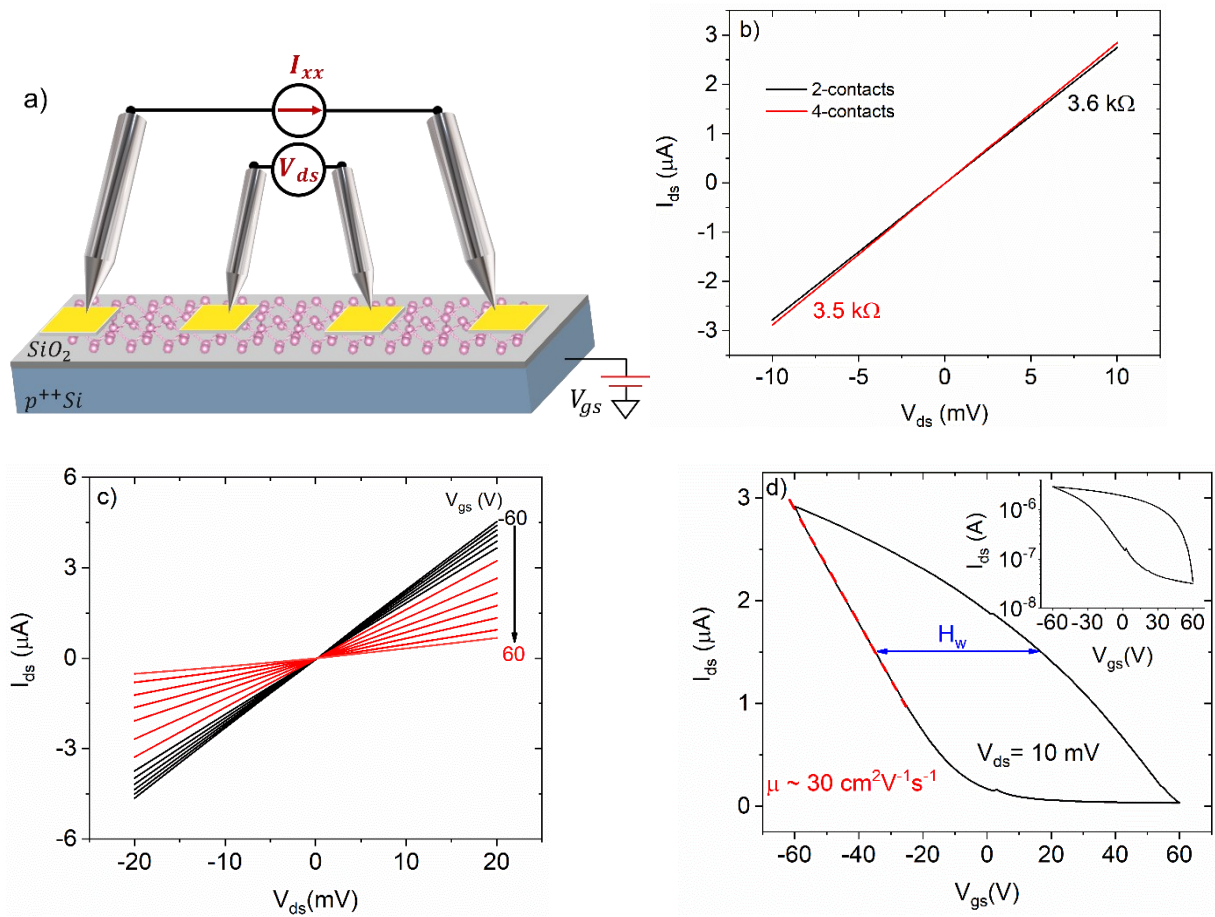

*Figure 2 – a) Schematic of the setup used for the four-probe electrical characterization. The current ($I_{xx}$) is forced between the outer contacts while the voltage drop ($V_{ds}$) is measured between the inner contacts. b) $I-V$ characteristics measured in two- and four- probe configuration. c) Output characteristic of the device recorded in two-probe configuration with the gate voltage ranging from $-60\ V$ to $60\ V$ (red lines are referred to positive gate voltages). d) Transfer characteristic measured over a loop of the gate voltage between -60 V and 60 V. The dashed red line represents the linear fit used to estimate the field effect mobility. The inset shows the transfer curve with the current on a logarithmic scale.*

The output characteristics, i.e. the drain-source current ($I_{ds}$) as a function of the voltage drop between two inner contacts ($V_{ds}$) with the gate-source voltage ($V_{gs}$) as control parameter, exhibit a linear behaviour. The application of a negative gate bias leads to an increase of the channel current, typical of a p-type device, and does not affect the linearity of the $I_{ds} - V_{ds}$ characteristics.[36] The transfer characteristic, i.e. the $I_{ds} - V_{gs}$ curve measured at fixed $V_{ds} = 10\ mV$ over a loop of the gate voltage is reported in Figure 2d, confirming the p-type behaviour, with a modulation of about two orders of magnitude. We did not fully reach the off state of the transistor in the applied voltage range, which was intentionally limited to avoid the breakdown of the gate dielectric. Considering the non-linear behaviour of the transfer characteristics, the channel current can be expressed as:

$$I_{ds} = \frac{W}{L}\mu_{FE}C_{ox}V_{ds}|V_{gs} - V_{th}|^{\alpha} \qquad (1)$$

where $W = 10\ \mu m$ and $L = 5\ \mu m$ are the channel width and length, respectively, $\mu_{FE}$ is the field-effect mobility, $C_{ox} = 3.84 \cdot 10^{-8}\ \frac{F}{cm^2}$ is the capacitance per unit area of the gate dielectric, $V_{th}$ is the threshold voltage and $\alpha \geq 1$ is a dimensionless parameter that accounts for a possible $V_{gs}$-dependence of the mobility[37]. According to Eq. (1), when the $I_{ds} - V_{gs}$ curve is linear, $\alpha = 1$, and the mobility can be obtained as:

$$\mu_{FE} = \frac{L}{W}\frac{1}{C_{ox}V_{ds}}\frac{dI_{ds}}{dV_{gs}} \qquad (2)$$

From the linear fit, we obtained a relative high $\mu_{FE} \sim 30\ cm^2V^{-1}s^{-1}$, considering that the sample was not subjected to any functionalization, that is slightly below both the theoretically predicted[38,39] and the experimentally measured mobilities in similar devices.[40,41] We note that such a mobility is higher or comparable with the one obtained in $SiO_2$ back-gate FETs fabricated with other 2D materials.[42–45] The large hysteresis width, $H_w \sim 60\ V$, here defined as the $V_{gs}$ difference corresponding to the current of $1.5\ \mu A$, is mainly due to charge trapping impurities and has already been reported for similar devices.[46–49] Gate-induced hysteresis can be ascribed to charge transfer from/to intrinsic and extrinsic trap states. Intrinsic traps correspond mostly to BP crystal defects such as phosphor vacancies,[50] or grain boundaries,[51] while extrinsic traps are related to environmental adsorbates, like water and oxygen molecules, defects located at the metal/BP interface, and at the BP/$SiO_2$ interface.[46]

Although oxidation of the surface involves the dissociative chemisorption of oxygen that causes the decomposition of BP and the decrease in FET performance, we expect that water and oxygen play a marginal role in the formation of hysteresis because the electrical measurements were carried out in high vacuum to remove the surface adsorbates. Thus, traps at the BP/$SiO_2$ interface, intrinsic BP defects and border traps in $SiO_2$ as well as mobile charge in the $SiO_2$ layer are most likely responsible for the device hysteresis.

To better understand the origin of hysteresis we measured the transfer characteristics at different sweeping times. Figure 3a shows several transfer characteristics with $H_w$ increasing as function of the voltage sweep duration in the range $12 - 322\ min$. The exponential fit, showed in Figure 3c, reveals a RC- growth with a single time constant $\tau = 191\ min$. Such a long characteristic time reduces the probability that BP/$SiO_2$ interface traps contribute to the hysteresis since it has already been reported that interface traps are fast traps.[52] Then, slow trap states can be ascribed to both BP and $SiO_2$ trap states. Particularly, slow border traps in $SiO_2$ have already been reported and attributed to trivalent silicon dangling bonds or hydrogenic defects.[53] From the time constant $\tau$, it is possible to evaluate the involved capacitance as $C = R/\tau$, where $R$ is the inverse of transconductance $g_m = \frac{\partial I_{ds}}{\partial V_{gs}} \sim 0.3\ nS$. We obtain $C \sim 3\mu F$, which is higher than gate oxide capacitance, in the order of the $pF$, implying that the trap-related capacitance is the dominant one.

Such a capacitance can also be obtained through the sub-threshold swing $SS$ that is the gate voltage change corresponding to one-decade increase of the transistor current. Indeed, $SS$ is a function of the charge trap capacitance and the depletion layer capacitance according to the following relation:

$$SS = \frac{dV_{gs}}{dLog(I_{ds})} \approx \log(10)\frac{kT}{q}\left(1 + \frac{C_T + C_{DL}}{C_{OX}}\right) \qquad (3)$$

where, $k$ is the Boltzmann constant, $T = 300\ K$ is the temperature and $q$ is the electric charge. By neglecting the depletion layer capacitance, $C_{DL}$, with respect to the charge trap capacitance, $C_T$, (a reasonable assumption considering the low modulation of the current) and, having obtained $SS \sim 30\ V/dec$, we estimated a trap capacitance, $C_T \sim 20\ \mu F$, that is consistent with the one previously obtained.

To corroborate the hypothesis that slow trap states contribute to hysteresis we measured the transfer characteristic while irradiating the device with a super-continuous white laser source at $50\ mW/mm^2$; after that, we repeated the measurements in dark every 10 minutes. Figure 3b shows that under illumination (black line) the device conductance and the hysteresis width are enhanced. This result is expected as the illumination causes both the generation of electron-hole pairs and excitation of trapped charges, which increase the carrier concentration in the material. Charged traps, emptied of free carriers, induce an enlargement of the hysteresis. By repeating the measurements in the dark, the conductance as well as the hysteresis decrease until they stabilize after about an hour. Figure 3d shows how $H_w$ varies after light irradiation. The best fit is obtained through a double decreasing exponential, one with a small time constant ($\sim 5\ s$) and a predominant one with a long characteristic time ($\sim 42\ min$). The fast time constant is related to electron-hole pair recombination, while the longer one supports the conclusion that hysteresis is dominated by intrinsic deep slow traps.

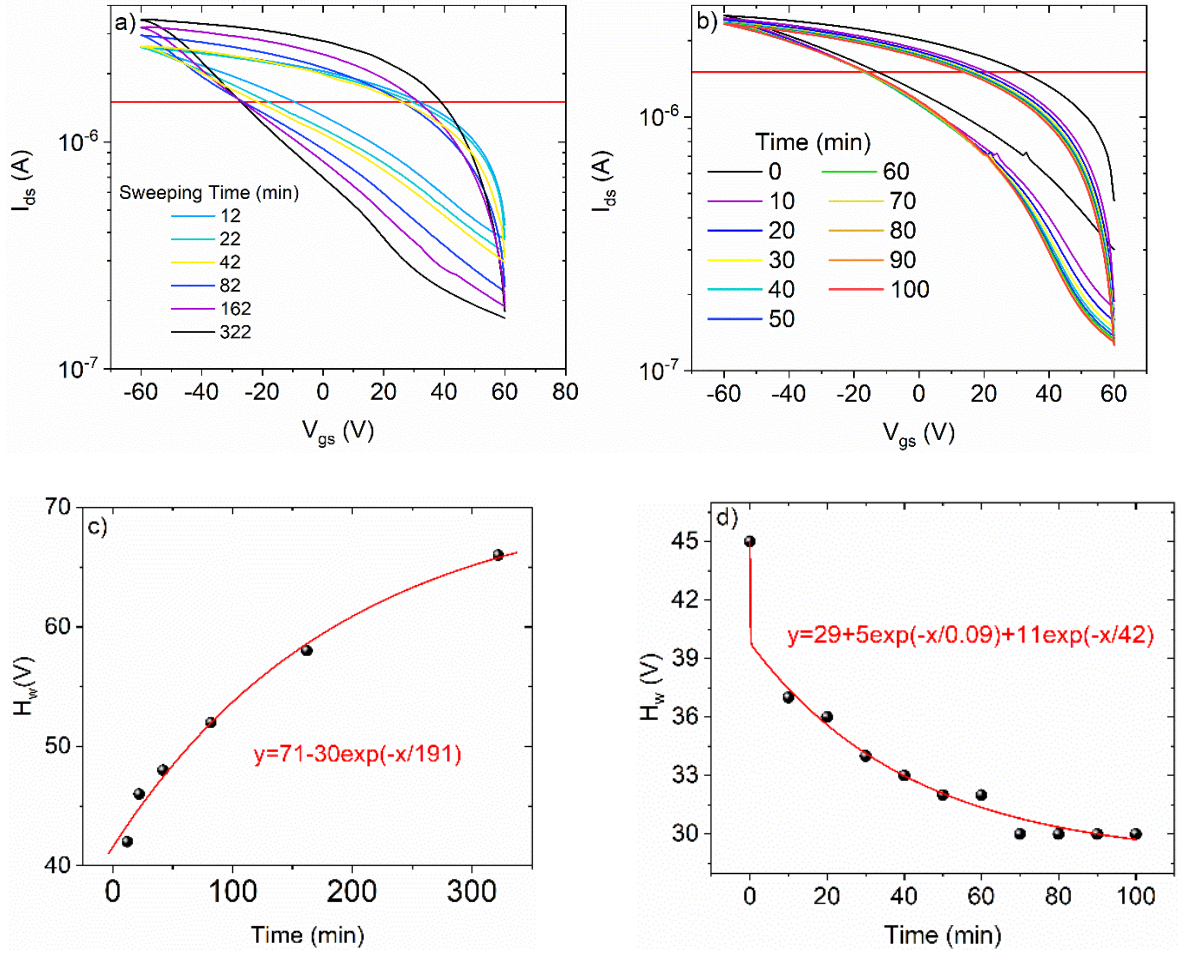

*Figure 3 – a) Transfer characteristics recorded at different sweeping times ranging from 12 min to 322 min. The horizontal red line indicates the current value used to evaluate the hysteresis width. b) Transfer characteristics measured under supercontinuous white laser illumination (black line) and in dark every 10 minutes after the laser was switched off. c) Exponential fit of the hysteresis width as function of the gate voltage sweeping time, revealing a time constant of* 191 *min. d) Double exponential fit of the hysteresis width obtained from figure b revealing a fast characteristic time of* 5 *s and a long one of* 42 *min.*

Although the presence of trap states is detrimental to FET performance, we here show that the charge trapping can be exploited to realize a memory device. We highlight that in such a way we realize non-volatile memory devices with good performance without using an additional charge-trapping layer, but just exploiting the presence of BP and $SiO_2$ defects.

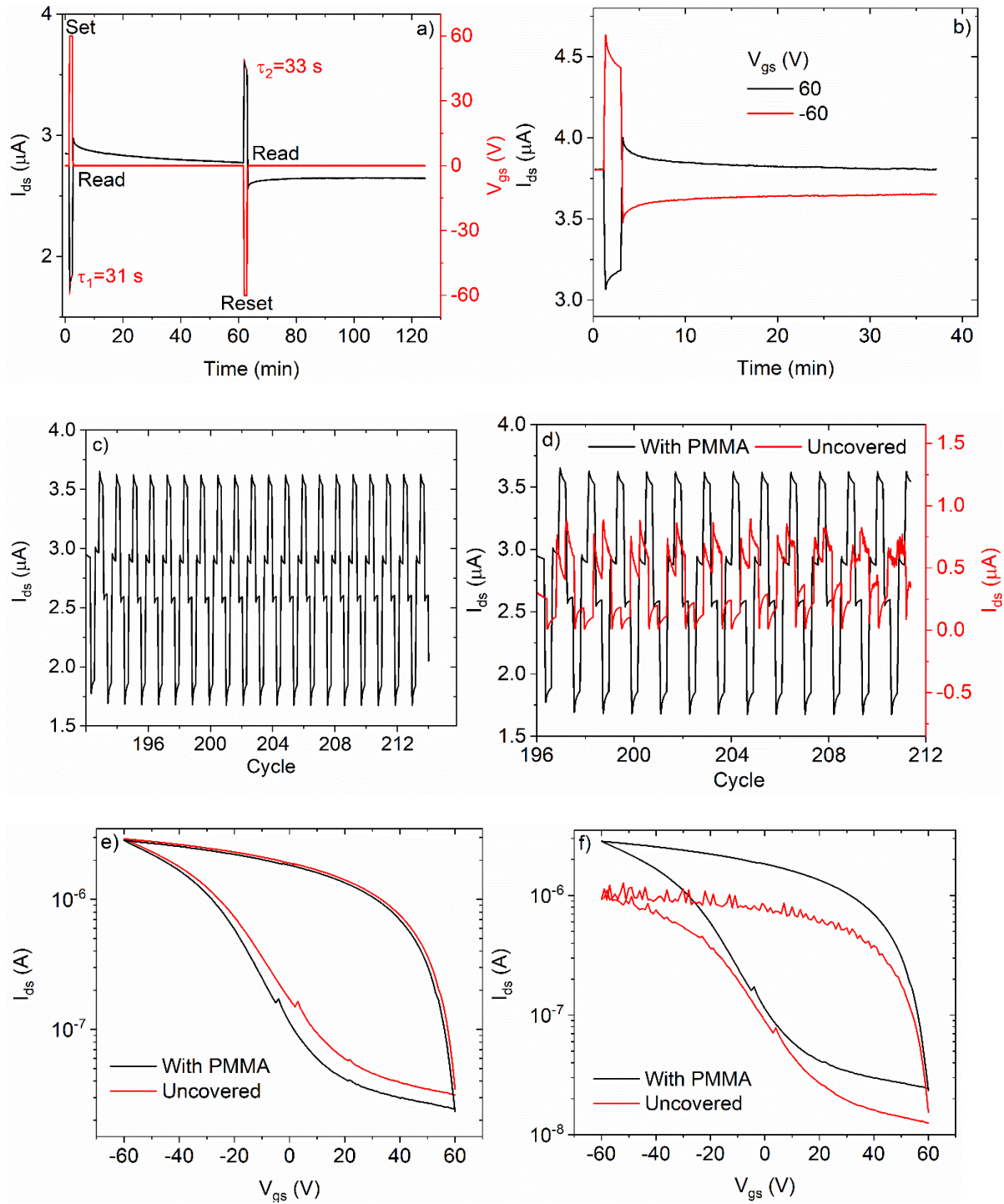

Figure 4 – a) Channel current (black line) recorded under gate pulses (red line) at $\pm 60\,V$ showing single set-read-reset-read cycles of the non-volatile memory. b) Transient behaviour of the channel current after the gate pulses at $\pm 60\,V$ revealing a retention time longer than 30 minutes. c) Repeated set-read-reset-read cycles of the memory device showing an endurance

*over 200 cycles. d) Repeated set-read-reset-read cycles measured after a month air exposure of a device covered with PMMA (black line) and an uncovered one (red line). e) Transfer characteristics of two BP FETs, one covered with PMMA (black line) and the other exposed to air (red line), recorded immediately after the device fabrication. f) Transfer characteristics of the same devices of e), measured after one-month exposure to air.*

Figure 4a shows the transient behaviour of the device, investigated through a series of $V_{gs} = \pm 60\ V$ pulses while the $I_{ds}$ current is monitored over time. While the gate pulse is in the high positive or negative state the channel current is not constant, but it increases/decreases in an exponential way. This is due to trapping/detrapping of charge inside the trap states. When the gate voltage is set to $0\ V$, the current tends to return to its initial value. However, the transient behaviour after a positive and a negative gate pulse shows that $I_{ds}$ stabilizes at two different values, distinct from the initial one (Figure 4b). This separation is retained for a time longer than 30 minutes, which is comparable with the retention time observed in non-volatile BP memories with charge trapping layers[20–22]. Then, we tested the endurance of the device by continuously applying pulses of 1 min width at $V_{gs} = \pm 60\ V$. Figure 4c shows that the device response is stable after 200 cycles, consistently with similar non-volatile memories realized with other 2D materials.[43]

Finally, since the main obstacle to the realization of BP-based devices is the lack of stability in the air, we covered several BP devices with a PMMA layer. Then, we performed endurance tests on two similar devices, with and without PMMA. Initially, the devices have similar behaviour but, after exposure to air for a month, the PMMA protected device maintains its current while the unprotected device shows evident signs of current deterioration (Figure 4d). Figure 4e compares the transfer characteristics of the PMMA covered and uncovered devices immediately after the fabrication process. The similarity of the two curves confirms that encapsulation by PMMA does not significantly alter the BP device. The transistor covered with PMMA has a slightly larger hysteresis because PMMA can contribute to charge trapping. Figure 4f shows the same transfer characteristics recorded after that both devices were left exposed to the air for a month. The device covered with PMMA does not show any signs of deterioration while the uncovered FET exhibits an evident reduction in conductivity and a much noisier current due to surface oxidation. This result confirms that PMMA encapsulation is a simple and effective method of avoiding the deterioration of air exposed BP memory FETs.

## 4. Conclusion

We reported the fabrication and the electrical characterization of BP field effect transistors. We exploited the presence of a hysteresis in the FET transfer characteristic to realize non-volatile memories. Gate voltage pulses allowed to store electric charges inside intrinsic BP and $SiO_2$ traps states. The obtained device showed an endurance over 200 cycles and retention longer than 30 minutes. Finally, encapsulating BP with a PMMA protective layer was used as a simple way to preserve the electrical properties of the memory over a month.

**Acknowledgements**
We thank the University of Salerno, Salerno, Italy, for the grants ORSA200207 and ORSA195727.

Received: ((will be filled in by the editorial staff))
Revised: ((will be filled in by the editorial staff))
Published online: ((will be filled in by the editorial staff))

TOC – Table of contents

The transfer characteristics of back-gated field effect transistors with black phosphorus channel exhibit wide hysteresis caused by intrinsic defects. Hysteresis is exploited to realize non-volatile memories with endurance over 200 cycles and retention longer than 30 minutes. Poly (methyl methacrylate) layer is used to slow down the degradation of the device in air.

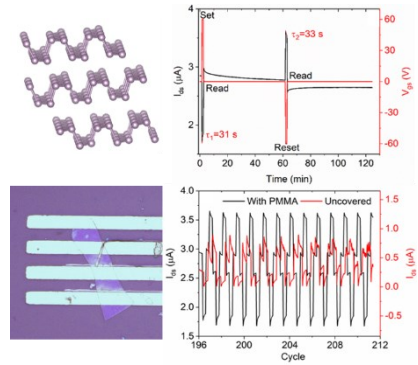